\documentclass{aa}  

\usepackage{graphicx}
\usepackage{txfonts}
\usepackage{float}
\usepackage{amsmath}
\usepackage{subfig}
\usepackage{multirow}
\usepackage{amssymb}
\usepackage{dirtytalk}
\usepackage{xcolor}
\usepackage{hyperref}
\usepackage{cleveref}
\crefformat{section}{\S#2#1#3} 
\crefformat{subsection}{\S#2#1#3}
\crefformat{subsubsection}{\S#2#1#3}
\begin{document}

 \title{Waking the monster: the onset of AGN feedback \\in galaxy clusters hosting young central radio galaxies}
\titlerunning{Waking the monster: the onset of AGN feedback}
   \author{F. Ubertosi
          \inst{1,2}
           \and
           M. Gitti
            \inst{1,3}
          \and
          F. Brighenti
            \inst{1,4}
            \and
            V. Olivares
            \inst{5}
            \and
            E. O'Sullivan
            \inst{6}
            \and
            G. Schellenberger
            \inst{6}
          }

   \institute{$^{1}$Dipartimento di Fisica e Astronomia, Università di Bologna, via Gobetti 93/2, I-40129 Bologna, Italy\\
              \email{francesco.ubertosi2@unibo.it} \\
     $^{2}$Istituto Nazionale di Astrofisica - Osservatorio di Astrofisica e Scienza dello Spazio (OAS), via Gobetti 101, I-40129 Bologna, Italy\\
     $^{3}$Istituto Nazionale di Astrofisica - Istituto di Radioastronomia (IRA), via Gobetti 101, I-40129 Bologna, Italy\\
     $^{4}$University of California Observatories/Lick Observatory, Department of Astronomy and Astrophysics, University of California, Santa Cruz, CA 95064, USA \\
     $^{5}$Department of Physics and Astronomy, University of Kentucky, 505 Rose Street, Lexington, KY 40506, USA \\
     $^{6}$Center for Astrophysics | Harvard \& Smithsonian, 60 Garden Street, Cambridge, MA 02138, USA
    }

   \date{Accepted for publication in A\&A}

 
  \abstract
    {}
    {The investigation of the feedback cycle in galaxy clusters has historically been performed for systems where feedback is ongoing (``mature-feedback'' clusters), that is where the central radio galaxy has inflated radio lobes, pushing aside the intracluster medium (ICM). In this pilot study, we present results from ``pre-feedback'' clusters, where the central newly active radio galaxies (age $<10^{3}$ yr) may not yet have had time to alter the thermodynamic state of the ICM.}
    {We analyze {\it Chandra} and MUSE observations of two such systems, evaluating the hot gas entropy and cooling time profiles, and characterizing the morphology and kinematics of the warm gas.}
    {Based on our exploratory study of these two sources, we find that the hot gas meets the expectations for an as of yet unheated ICM. Specifically, the entropy and cooling time of pre-feedback clusters within 20 kpc from the center fall below those of mature-feedback clusters by a factor $\sim$2. We speculate that with an estimated mechanical power of $\sim10^{44} - 10^{45}$ erg s$^{-1}$, the two young radio galaxies may restore the entropy levels in a few tens of millions of years, which are typical values of power outbursts and lifetimes for radio galaxies in clusters. Conversely, the properties of the gas at $\sim10^{4}$ K seem to remain invariant between the two feedback stages, possibly suggesting that the warm gas reservoir accumulates over long periods ($10^{7}$--$10^{8}$ yr) during the growth of the radio galaxy. 
    We conclude that the exploratory results obtained from our analysis of two cluster-central young radio galaxies are crucial in the context of understanding the onset of  active galactic nuclei  feedback, and they provide enough motivation for further investigation of similar cases.
    }
    {}


   \maketitle
%
\section{Introduction}\label{sec:intro}
The heating and cooling balance in galaxy clusters has historically been investigated in systems where feedback from active galactic nuclei (AGN) is currently regulating the thermodynamics of the intracluster medium (ICM; e.g., \citealt{2004ApJ...607..800B,2011ApJ...735...11O,2021Univ....7..142E,2022PhR...973....1D}). These studies have highlighted how the radio lobes of the central AGN, while expanding, push aside the cooling gas and excavate depressions in the ICM. \\Over the last twenty years, it has emerged that the gas entropy and cooling time are sensitive proxies of cooling regulation by AGN activity. On the one hand, systems with entropy $\leq$30 keV cm$^{2}$ and a cooling time $\leq$0.5--1 Gyr within the central tens of kiloparsecs typically host filamentary warm and cold gas likely originating from ICM condensation (e.g., \citealt{2016ApJ...830...79M,2019A&A...631A..22O,2022A&A...666A..94O,2022ApJ...928..150T,2022ApJ...940..140C}). On the other hand, ICM entropy profiles are powerful indicators of the impact of feedback: an excess at $\sim$10--30 keV~cm$^{2}$ in the inner $\sim$20--30 kpc with respect to the inward extrapolation of the outer profile has been interpreted as the result of energy injection due to AGN activity (e.g., \citealt{2008ApJ...687..899R,2009ApJS..182...12C,2009A&A...501..835M,2013ApJ...776...84C,2020ApJ...905...50P,2022MNRAS.515.4838N}).
\\ However, the above picture was drawn from systems where feedback is already ongoing, whereas the conditions that lead to its onset are currently unknown. A further step is to investigate systems that are right around the point of triggering feedback, as soon after the AGN jets start up as possible. 
\\ A class of radio galaxies meeting this requirement is that of young AGN, extended on galactic or subgalactic scales. Their distinctive feature is the peaked radio spectrum: if the peak is found at frequencies $\lessapprox$ 100 MHz, the source is classified as a compact steep spectrum (CSS) radio galaxy (extended for a few kiloparsecs and typically $\leq$10$^{5}$ yr old); a peak around GHz frequencies identifies a gigahertz-peaked spectrum (GPS) source, with a largest linear size of less than 1 kpc and an age of $\leq$10$^{4}$ yr (see e.g., \citealt{1998PASP..110..493O,2016AN....337....9O,sadler2016,2021A&ARv..29....3O}). \\Given such short timescales and the fact that studies of AGN feedback in galaxy clusters typically target extended radio galaxies, there is a paucity of brightest cluster galaxies (BCGs) in cool-core clusters with known GPS- or CSS-stage AGN. An example is given by the CSS source 1321+045, which was recently studied by \citet{2021ApJ...913..105O}. The authors found that despite having a low central entropy and cooling time (9 keV cm$^{2}$ and 390 Myr within 8 kpc), the overall properties of the host cluster are similar to those of other objects with extended central AGN. However, the CSS source 1321+045, having a size of 16 kpc, might have already influenced the surrounding ICM. \\This work presents our investigation of the onset of feedback in even younger, smaller sources, the GPS radio galaxies. In the following we assume a $\Lambda$CDM cosmology with H$_{0}$=70 km s$^{-1}$ Mpc$^{-1}$, $\Omega_{\text{m}}$=0.3, and $\Omega_{\Lambda}$=0.7. 
   \begin{figure*}[ht]
   \centering
   \includegraphics[width=\linewidth]{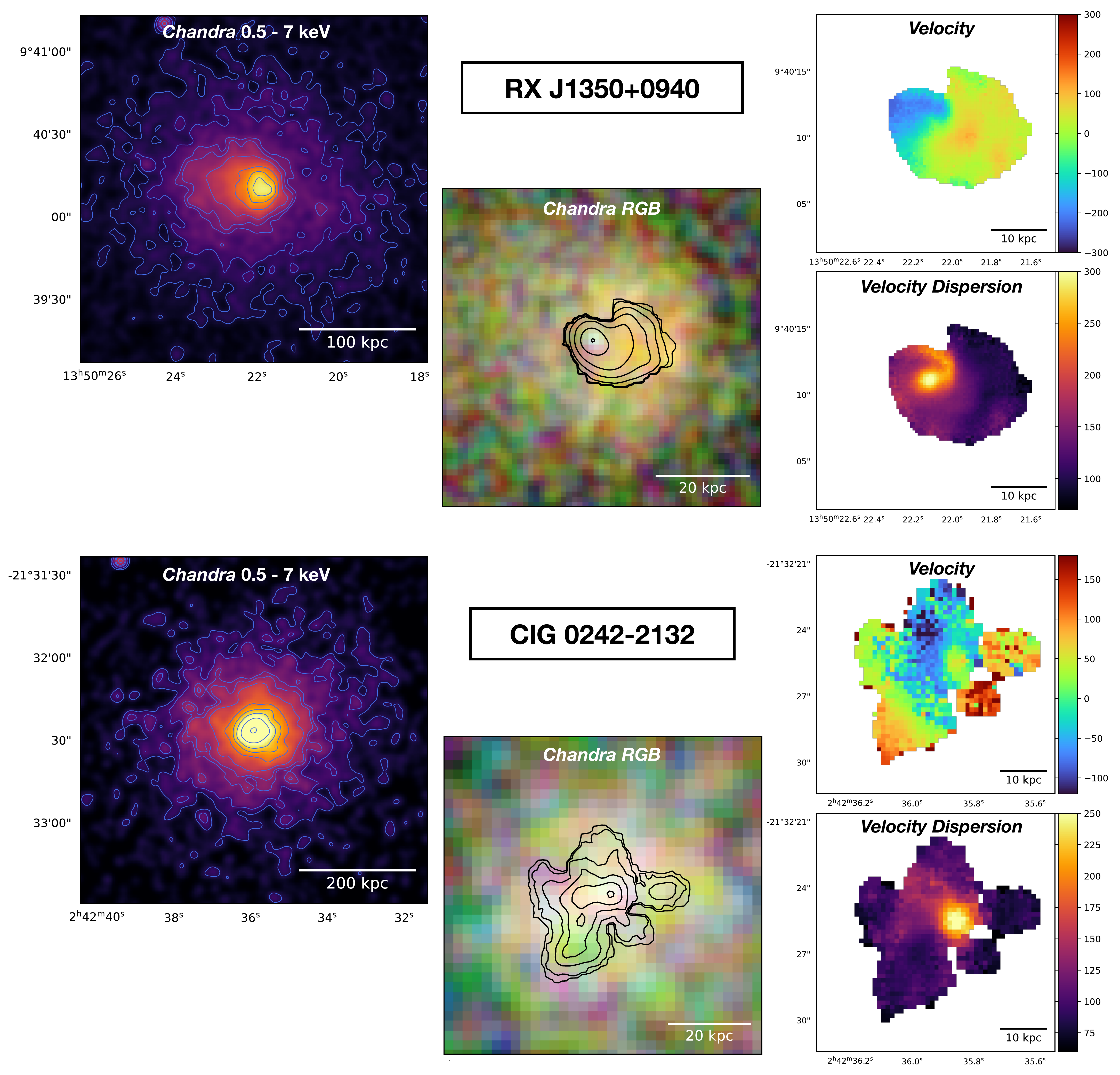}
      \caption{{\it Chandra} and MUSE images of the targets. For each object, we show in the \textit{left panel} a (background-subtracted
and exposure-corrected) \textit{Chandra} image  in the 0.5--7 keV band of the large-scale ICM emission. Contours start from the peak of the X-ray emission and decrease in steps by a  factor of two. The images have been smoothed with a Gaussian of $\sigma=3$ pixels. The \textit{middle panel} shows a RGB \textit{Chandra} image of the central region of each cluster, with the bands 0.3--1.2, 1.2--2.5, and 2.5--7.0 keV being shown as red (R), green (G), and blue (B), respectively. Overlaid in black are the total intensity H$\alpha$ contours from the MUSE data. In the \textit{right panels,} we show the warm gas kinematics (top) and velocity dispersion (bottom).
              }
         \label{imgcluster}
   \end{figure*}

\section{Selection of objects}\label{sec:selection}
\vspace{-0.1cm}
We selected targets from the sample of \citet{2015MNRAS.453.1201H,2015MNRAS.453.1223H} by choosing objects with a peaked radio spectrum, typical of GPS sources. Furthermore, we excluded sources with additional power-law components at MHz frequencies to avoid restarted sources (i.e., young AGN embedded in large-scale older radio emission, see e.g., \citealt{2021Galax...9...88M}). This selection returns a list of only five systems (Tab. \ref{tab:objects}). 
\\Out of the objects listed in Tab. \ref{tab:objects}, in this work we focus on RX~J1350+0940 and ClG~J0242-2132, which are the only ones with deep enough X-ray (\textit{Chandra}) and optical -- Multi-Unit Spectroscopic Explorer (MUSE) -- data to assess the hot and warm gas properties in the inner 20 kpc, where cooling and feedback are typically at play. In \cref{app:comments}, we summarize the data reduction for the {\it Chandra} and MUSE observations. We stress that the results presented here for these two clusters represent only a preliminary attempt to determine the thermodynamic state of clusters before the onset of feedback. A larger sample of similar systems would be required to provide definitive results.
\\ The AGN in the two clusters have the typical radio spectrum of GPS sources, peaking at a rest-frame frequency of $\nu_{p} = 3.9$ GHz and of $\nu_{p} = 0.8$ GHz, respectively \citep{2015MNRAS.453.1223H}. 
Both radio galaxies in RX~J1350+0940 and ClG~J0242-2132 are unresolved on arcsecond (kiloparsec) scales (\citealt{2020ApJS..247...53K} and \citealt{2015MNRAS.453.1223H}, respectively), and reveal small extension only on milliarcsecond (parsec) scales \citep{2015MNRAS.453.1223H}. The nondetection of extended emission at MHz frequencies implies -- assuming a typical magnetic field of a few $\mu$G (e.g., \citealt{2004IJMPD..13.1549G}) -- a timescale of at least $10^{8}$ yr since the last radio activity (e.g., \citealt{2017A&A...600A..65S}). Information on the parsec-scale properties of both sources are available from the VLBA Calibrator List Tool \footnote{\url{https://obs.vlba.nrao.edu/cst/}; \citealt{2020A&A...644A.159C}.}: at 2.3 GHz, the radio galaxy in RX~J1350+0940 has a largest linear size (LLS) of 90 pc, while that in ClG~J0242-2132 has a LLS of 200 pc. Assuming an expansion speed of $0.2\,c$ (e.g., \citealt{2009AN....330..193G}), the implied kinematic ages are $t_{kin}\sim700$ yr and $t_{kin}\sim1600$ yr, respectively (see Tab. \ref{tab:objects}). 
\\ To identify any peculiarity of these \say{pre-feedback} clusters, we need to compare our results with the general population of \say{mature-feedback} clusters. We use the pre-feedback term to describe the systems where the central radio galaxy extends on subkiloparsec scales and is young enough so that no cavities or shocks could have impacted the ICM, while we use the mature-feedback term to describe systems where the kiloparsec-scale lobes of the central AGN extend into the ICM, pushing aside the gas. 
We adopt the Archive of $Chandra$ Cluster Entropy Profile Tables (ACCEPT) sample of \citet{2009ApJS..182...12C} as representative of the average properties of mature-feedback systems. 
To restrict the comparison to systems that are comparable to RX~J1350+0940 and ClG~J0242-2132 in terms of mass and dynamical state, we selected the ACCEPT systems with an average temperature $\geq$2 keV (thus avoiding galaxy groups and elliptical galaxies, e.g., \citealt{2021Univ....7..139L}) and a central entropy $\leq$30 keV cm$^{2}$ (to consider only objects with a cool core).  

   \begin{figure*}[ht]
   \centering
   \includegraphics[width=\linewidth]{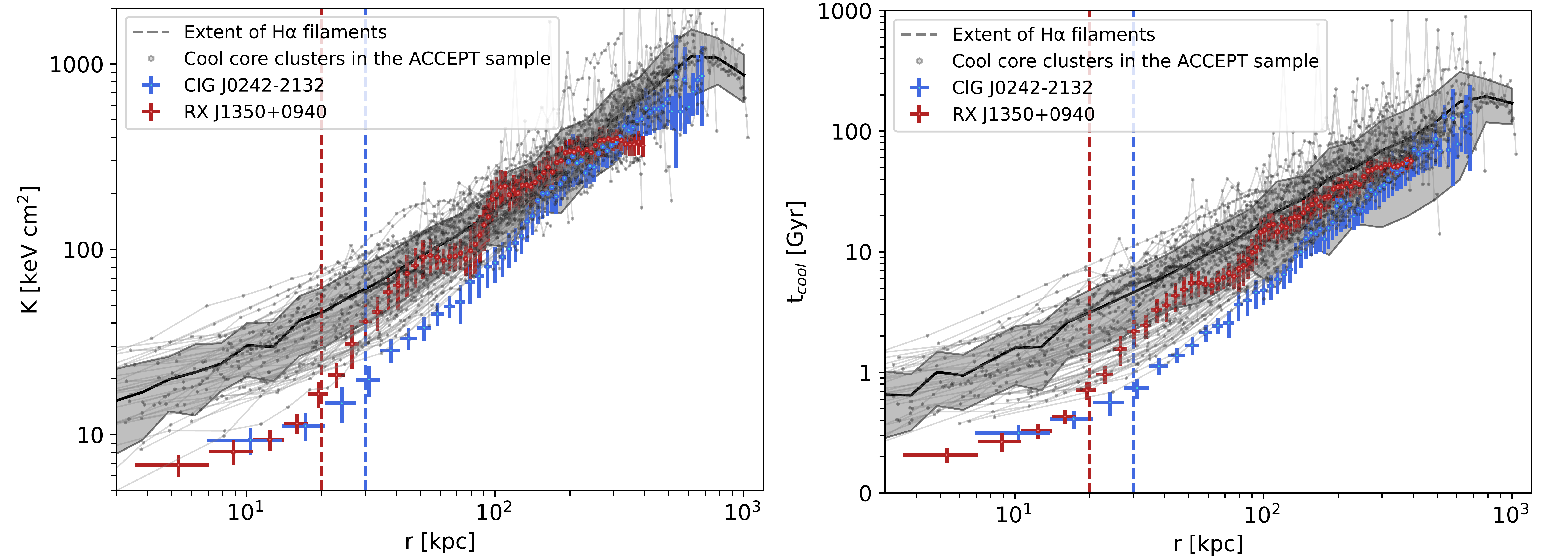}
      \caption{Profiles of ICM entropy (left) and cooling time (right) for the ACCEPT cool-core clusters with $kT \geq 2$ keV are plotted in gray, while those of ClG~J0242-2132 and RX~J1350+0940 are plotted in blue and red, respectively. The black line and the shaded gray area represent the average profile and the scatter of the ACCEPT clusters, respectively. Readers can refer to \cref{subsec:profiles} for details.
              }
         \label{profiles}
   \end{figure*}
   
\section{Results}\label{sec:results}
\subsection{ICM morphology and warm gas kinematics}\label{subsec:morphology}
We show in Fig. \ref{imgcluster} a multiscale, multiwavelength view of the two objects. In RX~J1350+0940, the {\it Chandra} image shows an asymmetric ICM distribution, with a surface brightness edge west of the center. The RGB image in the middle panel confirms this asymmetry and reveals that the cool ICM is primarily located west of the central AGN. The MUSE H$\alpha$ contours (total luminosity L$_{\text{H}{\alpha}} = 9.6\times10^{41}$~erg~s$^{-1}$) show an extended spiral morphology (with LLS$\sim$20 kpc), cospatial with the X-ray brightest region of the ICM. These structures suggest that the hot and warm gas distributions have been subject to sloshing (see e.g., \citealt{kokotanekov2018}). The velocity structure of optical-emitting gas of RX~J1350+0940 reveals a clear gradient from $-$200 to $+$ 30~km~s$^{-1}$ from the northwest to northeast of the central galaxy. The velocity dispersion,  $\sigma_{\rm gas}$, peaks with 300~km~s$^{-1}$ at the position of the BCG and decreases from 250~km~s$^{-1}$ to 90~km~s$^{-1}$ following the sloshing spiral.
The cluster ClG~J0242-2132 shows a rather symmetric morphology from the large-scale \textit{Chandra} image and the RGB zoom-in. The 0.3 -- 1.2 keV ICM is evenly distributed around the central AGN, and the warm gas phase traced by the H$\alpha$ contours (L$_{\text{H}{\alpha}} = 1.7\times10^{42}$~erg~s$^{-1}$) has a filamentary structure extending roughly equally in all directions. The optical data reveal several radial filaments across the whole azimuth, with LLS$\sim$40 kpc and a chaotic velocity field. The velocity dispersion is high, $>$120~km~s$^{-1}$, in the region to the northwest of the central BCG, likely due to several unresolved filaments overlapping along the line of sight. The rest of the filaments show velocity dispersions on the order of 80~km~s$^{-1}$.
\\ Overall, the extent and morphology of the warm gas in the two objects are similar to those of filaments in mature-feedback BCGs (see e.g., \citealt{2018ApJ...865...13T,2019A&A...631A..22O,2021A&A...649A..23C}). The similarity is also evident from the gas kinematics. In particular, the cluster ClG~J0242-2132 has an average $\sigma_{\rm v}$ ($\sim$100~km~s$^{-1}$, excluding the inner 2 kpc) that is similar to that of ten BCGs in mature-feedback clusters observed with MUSE, where $\langle \sigma_{\rm v} \rangle\sim$ 110~km~s$^{-1}$ \citep{2019A&A...631A..22O}. RX~J1350+0940 has a slightly larger velocity dispersion along the filaments of 130~km~s$^{-1}$, but it is still close to the average of the ten BCGs. 
\subsection{ICM entropy and cooling time profiles}\label{subsec:profiles}
To test the efficiency of ICM cooling, we derived radial profiles of entropy, $K$, and a cooling time, $t_{\text{cool}}$. These two quantities are a function of the ICM temperature, $kT$, and electron density, $n_{e}$, via the following:
\begin{equation}
    K = \frac{kT}{n_{e}^{2/3}}
    \label{entro}
\end{equation}
\begin{equation}
    t_{\text{cool}} = \frac{\gamma}{\gamma -1} \frac{kT}{\mu \,X \,n_{\text{e}}\,\Lambda(\text{T, Z})}\,\text{,}
    \label{tcool}
\end{equation}
where $\gamma = 5/3$ is the adiabatic index, $\mu\approx0.6$ is the mean molecular weight, $X\approx0.7$ is the hydrogen mass fraction, and $\Lambda(\text{T,Z})$ is the cooling function (from \citealt{1993ApJS...88..253S}). To be consistent with the ACCEPT sample \citep{2009ApJS..182...12C}, the electron density was measured from a radial surface brightness profile (see \cref{app:alternative}).
The temperature was determined by fitting a \texttt{projct$\ast$tbabs$\ast$apec} model to the spectrum of concentric annuli (centered on the AGN coordinates, see Tab. \ref{tab:objects}) with at least 1000 counts in each bin. The central 1.5$''$ were excised to avoid contamination from the nuclear X-ray point source found in both clusters. The temperature profile was interpolated on the grid of the density profile to obtain the gas entropy and cooling time profiles in 1.5$''$-- wide annular rings.
Testing other methods to derive these quantities provided consistent results (see \cref{app:alternative}). In Fig. \ref{profiles},  we show the entropy and cooling time radial profiles for ClG~J0242-2132 and RX~J1350+0940, and we over-plotted the profiles of the ACCEPT cool-core clusters, the average ACCEPT profile, and its scatter (see \cref{sec:selection}; \citealt{2009ApJS..182...12C}). 
\\At large radii, the profiles of both clusters are consistent with the ACCEPT sample. However, in the inner few tens of kiloparsecs, the profiles deviate from mature-feedback clusters, showing lower entropy and cooling time (see Fig. \ref{profiles}). For both clusters, the extent of the H$\alpha$ filaments approximately traces the region where such a deviation occurs (vertical lines in Fig. \ref{profiles}). The average entropy of mature-feedback clusters in the inner 20 kpc is $K^{ACC} = 28.5\pm14.0$ keV cm$^{2}$, while that of the two objects in our sample is $K^{GPS} = 10.4\pm1.9$ keV cm$^{2}$ (the associated uncertainty is the scatter around the mean). The average cooling time of mature-feedback clusters for $r\leq20$ kpc is $t_{cool}^{ACC} = 1.6\pm1.0$ Gyr, while that of our two clusters is $t_{cool}^{GPS} = 0.4\pm0.2$ Gyr. We attribute the undulation in the profiles of RX~J1350+0940 to sloshing of the ICM (see also \cref{subsec:morphology}). The profile of ClG~J0242-2132 is smooth between 8 kpc to 500 kpc, and it starts to deviate from the ACCEPT clusters at approximately 100 kpc from the center. We also note that RX~J1350+0940 and ClG~J0242-2132 have central cooling times of approximately 200 -- 300 Myr at $r \sim 10$ kpc, which is on the order of the typical time that has passed since the last episode of AGN activity (see \cref{sec:selection}).
\\We note that there are two ACCEPT clusters with entropy profiles that resemble those of the two pre-feedback clusters in the inner 10s of kiloparsecs. The one whose entropy decreases around 10 kpc and crosses the profile of RX J1350+0940 at $\sim$4 kpc is Abell 1991, whose central radio galaxy has been classified as a CSS radio source (see \citealt{2014PhDT.......338H}). Thus, this source may potentially be similar to the CSS~1321+045 studied in \citet{2021ApJ...913..105O}, and its decreasing profile may support our results on young radio sources in galaxy clusters. The other galaxy cluster is 2A0335+096 (e.g., \citealt{2009MNRAS.396.1449S}). We note that its profile is similar to those of the two young sources from large radii to $\sim$10 kpc, but it seems constant between 1 -- 10 kpc. Additionally, we point out that the cooling time profiles show a more marked difference, with the two pre-feedback clusters having the shortest cooling times within 20 -- 30 kpc from the center.
\\ The energy required to boost the entropy of the two clusters to the values of the mature-feedback objects can be measured as $\Delta E = M_{\text{gas}}\,\Delta Q$, where $M_{\text{gas}}$ is the gas mass within 20 kpc from the center and $\Delta Q$ is defined as follows (see \citealt{2012ApJ...759...87C}):
\begin{equation}
    \Delta Q = \frac{kT\,(<\text{r})}{(\gamma-1)\,\mu\,m_{p}}\,\left( \frac{K^{ACC} - K^{GPS}}{K^{ACC}} \right)
.\end{equation}
We derived $M_{\text{gas}}\,(\leq 20\,\text{kpc})$ by integrating the density profile over the spherical shells (see e.g., \citealt{2006MNRAS.368..518V}), finding $M_{\text{gas}}(\leq 20\,\text{kpc})\approx5\times10^{10}$ M$_{\odot}$ for RX~J1350+0940 and $M_{\text{gas}}(\leq 20\,\text{kpc})\approx10^{11}$ M$_{\odot}$ for ClG~J0242-2132. With $kT$($<$20 kpc) of 2.1 keV and 2.6 keV for RX~J1350+0940 and ClG~J0242-2132, we find that an energy of $\Delta E \sim 4\times10^{59}$ erg and $\Delta E \sim 10^{60}$ erg, respectively, is required to boost the entropy to the values of the mature-feedback sample. These are comparable to the typical outburst energy associated with X-ray cavities and shocks (e.g., \citealt{2006ApJ...652..216R}). 
\\To determine if the AGN could supply these energies, we followed \citet{2020ApJ...892..116W} to estimate the jet power, $P_{j}$, of the two young radio galaxies. Using the linear size, age (see Tab. \ref{tab:objects}), and radio luminosity at 5 GHz of the AGN ($9.1\times10^{41}$ erg s$^{-1}$ for RX~J1350+0940 and $1.3\times10^{43}$ erg s$^{-1}$ for ClG~J0242-2132), we find $P_{j} \sim 2\times10^{44}$ erg s$^{-1}$ for RX~J1350+0940 and $P_{j} \sim 10^{45}$ erg s$^{-1}$ for ClG~J0242-2132. For comparison, using the mechanical power versus 1.4 GHz radio luminosity relation derived by \citet{2011ApJ...735...11O}, we obtain $P_{j} \sim 3\times10^{44}$ erg s$^{-1}$ and $P_{j} \sim 2\times10^{45}$ erg s$^{-1}$, respectively, which are in agreement within a factor $\leq$2 from the previous estimate. As a note of caution, we observe that the two young sources, once grown to large sizes, may have a mechanical power that differs from these estimates, as the jet may vary in power depending on the surrounding medium that it crosses (e.g., \citealt{2011MNRAS.410.1527H,2018MNRAS.475.3493B}).
\\To understand the balance between heating and cooling, we must also take into account the ICM radiative losses over the same volume. These can be approximated by the bolometric X-ray luminosity of the two clusters, that is $L_{X}\,(<\text{20 kpc})=1.4\times10^{43}$ erg s$^{-1}$ for RX~J1350+0940 and $L_{X}\,(<\text{20 kpc})=1.5\times10^{44}$ erg s$^{-1}$ for ClG~J0242-2132. Thus, with a net power $P'_{j} = P_{j} - L_{X}$, and assuming that $P_{j}$ remains constant in time, we conclude that the AGN could raise the central entropy and balance the ICM radiative losses by supplying the $\Delta E$ estimated above in 35 - 70 Myr, which are typical lifetimes for radio galaxies \citep{2021Galax...9...88M}. 
\\Ultimately, to offer a possible comparison with simulations of AGN feedback, we observe that the fractional entropy difference within 20 kpc of $\frac{K^{ACC} - K^{GPS}}{K^{GPS}} = 1.7$ is consistent with the results of \citet{2012ApJ...746...94G} that the fractional amplitude of entropy fluctuations through the different stages is $\sim$2. We caution that this comparison is speculative, as the two pre-feedback clusters may not be perfectly described by the initial conditions in the simulations of \citet{2012ApJ...746...94G}. 
\section{Discussion}\label{sec:discussion}
The X-ray analysis of the two clusters with central newly active AGN ($\sim$10$^{3}$ yr) revealed that ICM entropy and cooling time in the innermost tens of kiloparsecs are different (by a factor of 2--3) from those of mature-feedback ACCEPT clusters (see \cref{sec:results}). For both clusters, the mechanical power of the young AGN seems tuned to match the magnitude of this difference, thus lifting the core entropy by a factor of 2 in a few tens of millions of years. Interestingly, such a dichotomy in entropy and cooling time was not found in the $\approx2$ Myr old CSS source 1321+04 \citep{2021ApJ...913..105O}. 
It is then possible that ICM heating in that cluster may have occurred rapidly, within the few million years required for the radio galaxy to grow to a size $\gtrapprox$10 kpc.
\\By contrast, the warm gas phase does not show any evolution between different stages of the feedback cycle. 
The morphological and spectral properties of the ICM in the two clusters considered here suggest that the warm gas is condensing out of the hot phase. According to \citet{2018ApJ...854..167G}, this occurs when the C ratio, $C = t_{cool}/t_{eddy}$ (where the eddy timescale $t_{eddy} = 2\pi\times(r^{2/3}L^{1/3})/\sigma_{\text{3D}}$ is the time a vortex requires to gyrate), is close to 1. 
Assuming an injection scale L = 20 kpc (the extent of H$\alpha$ filaments, see \citealt{2018ApJ...854..167G,2022A&A...666A..94O}) and $\sigma_{\text{3D}}\sim\sqrt{3}\,\langle\sigma_{v}\rangle$, we measured $C\sim1.5$ at r = 10 kpc. This indicates that the criterion for the formation of multiphase gas is met.
Therefore, the hot and warm phases seem related, given their similar morphology (see \cref{subsec:morphology}) and thermodynamic link. On the other hand, the different evolution of the two phases may suggest that the warm gas tank could have been inherited from a previous cycle of cooling and heating (at least $10^{8}$ yr ago, see \cref{sec:selection}). If the filaments were old (the typical survival time may be a few $10^{8}$ yr, e.g., \citealt{2022ApJ...924...24F}), we would expect turbulence to have been partially dissipated, hence a low $\sigma_{v}$ ($\leq$50--100~km~s$^{-1}$). This is the contrary of what we measured, suggesting that either turbulence is dissipated on longer timescales, or that other mechanisms (besides AGN feedback) can stir the multiphase gas (e.g., sloshing in RX~J1350+0940). Overall, it is possible that a residual tank of relatively turbulent, centrally concentrated warm gas is always present, and that new, external filaments (where $\sigma_{v}$ decreases) condense out of the ICM at each cooling episode. Such a scenario is consistent with the simulations of  \citet{2015ApJ...811...73L}, for example. 
\\ Based on our exploratory results on RX~J1350+0940 and ClG~J0242-2132, we hypothesize that at each feedback cycle the onset of heating may proceed as follows (see for comparison e.g., \citealt{2011MNRAS.411..349G,2015ApJ...811...73L,2022arXiv221111771W}). At the end of the last major episode of AGN activity, the ICM starts cooling again, and over a few $10^{8}$ yr the entropy and cooling time of the gas in the inner tens of kiloparsecs decrease to roughly half the values measured in clusters where feedback is present. The ICM condenses into a network of filaments. Multiphase gas fuels the AGN, which drives jets with high enough mechanical power to increase the central ICM entropy. After $\sim10^{3}$ yr (the current state of RX~J1350+0940 and ClG~J0242-2132), the ICM still has relatively low entropy and cooling time, as the radio galaxy has not yet deposited energy outside $r\sim1$ kpc, and condensation of the ICM into warm gas proceeds. In a few $10^{7}$ yr (but possibly as low as a few $10^{6}$ yr, as in the CSS source 1321+04, \citealt{2021ApJ...913..105O}), the entropy of the core has increased to the values observed in a mature-feedback cluster. The warm gas filaments are still observable as the ICM continues to replenish the warm gas tank, with mechanical uplift by radio lobe expansion likely stimulating further ICM condensation (e.g., \citealt{2008A&A...477L..33R,2015ApJ...802..118B}). These hypotheses may be confirmed by undertaking a similar analysis of a larger sample of other pre-feedback galaxy clusters.

\section{Summary \& conclusions}
We have presented an exploratory study on the onset of AGN feedback in galaxy clusters. We used {\it Chandra} and MUSE observations to investigate the ICM and warm gas properties in two clusters (ClG~J0242-2132 and RX~J1350+0940) hosting young central radio galaxies (age $\sim$10$^{3}$ yr) classified as GPS sources. Here we summarize our results:
   \begin{enumerate}
      \item The average ICM entropy and cooling time in the inner 20 kpc of pre-feedback clusters (10.4$\pm$1.9 keV cm$^{-2}$, 0.4$\pm$0.2 Gyr) are lower than in mature-feedback systems (28.5$\pm$14.0 keV cm$^{-2}$, 1.6$\pm$1.0 Gyr). The entropy of the former could be boosted to match that of the latter by injecting an energy of $\Delta E\approx10^{59-60}$ erg. Considering our tentative estimate of the mechanical power of the two young radio galaxies ($10^{44}$ -- $10^{45}$ erg s$^{-1}$), such energy injection may be achieved in a few tens of millions of years.
      \item The MUSE data reveal warm gas surrounding the BCG and extended for 20 - 30 kpc in radius, likely condensing from the cluster gas. In terms of spatial extent and kinematics, the line-emitting gas in pre-feedback clusters is similar to that observed in mature-feedback objects (the average $\sigma_{v}$ is $\sim$100 km~s$^{-1}$ in ClG~J0242-2132, $\sim$130 km~s$^{-1}$ in RX~J1350+0940, and $\sim$110 km~s$^{-1}$ in ten mature-feedback clusters, see \cref{subsec:morphology}).
      \item Altogether, these results point to ClG~J0242-2132 and RX~J1350+0940 being possible examples of strongly cooling clusters in which the activity of the young central radio galaxy has not yet affected the ICM and where the multiphase gas reservoir is accumulated while the radio galaxy grows to larger sizes. 
   \end{enumerate}
This work is the first observational attempt at characterizing the onset of feedback in galaxy clusters. To confirm our preliminary results, it is essential to undertake a similar investigation of a larger number of sources (such as those in Tab. \ref{tab:objects}). In the future, we also plan to investigate other probes of the multiphase gas, as well as the source of ionization of the warm phase, to build an evolutionary picture of AGN feedback in galaxy clusters.
\begin{acknowledgements}
We thank the anonymous reviewer for their useful suggestions, that improved our work. This research has made use of data obtained from the Chandra Data Archive and the Chandra Source Catalog, and software provided by the Chandra X-ray Center (CXC) in the application packages CIAO and Sherpa. Support for this work was provided by the National Aeronautics and Space Administration through Chandra Awards Number GO1-22125X and GO0-21112X issued by the Chandra X-ray Center, which is operated by the Smithsonian Astrophysical Observatory for and on behalf of the National Aeronautics Space Administration under contract NAS8-03060. V.O. was supported by NSF grant 2107711, Chandra X-ray Observatory grant GO1-22126X, and NASA grant 80NSSC21K0714. This research is based on observations collected at the European Southern Observatory under ESO programme(s) 0104.A-0801 and 0100.A-079.
\end{acknowledgements}

%
\bibliographystyle{aa} 
\bibliography{45894corr} 
%

\begin{appendix}
\section{Properties of the sources and data reduction}\label{app:comments}
In Tab. \ref{tab:objects}, we list the five GPS radio galaxies in cool-core clusters we selected from the sample of \citet{2015MNRAS.453.1201H,2015MNRAS.453.1223H} (see \cref{sec:selection}). The two objects in bold have deep enough {\it Chandra} and MUSE data to study the hot ICM and the warm gas, respectively, and to characterize their interplay. Regarding the {\it Chandra} data, RX~J1350+0940 has been observed for 20 ks (ObsID 14021), while ClG~J0242-2132 has been observed for 12 ks (ObsID 3266). The data have been reprocessed with CIAO-4.14 and CALDB-4.9.7, using standard data reduction\footnote{See \url{https://cxc.cfa.harvard.edu/ciao/threads/}.}. Point sources were masked during the analysis. Periods of background flaring were excluded from the data, and scaled blank-sky event files were used to create background spectra. These two objects have been observed with the Very Large Telescope using the MUSE integral-field spectrograph (IDs 0104.A-0801 and 0100.A-0792, for RX~J1350+0940 and ClG~J0242-2132, respectively). The MUSE data were reduced using the MUSE pipeline 2.8.5 \citep{2014ASPC..485..451W} and the EsoRex command-line tool, to obtain information on the warm gas component surrounding the BCG. We fit the data following the same method described by \citet{2019A&A...631A..22O}. The average seeing is 1.5$''$ for RX~J1350+0940 and 0.6$''$ for ClG~J0242-2132.
\begin{table*}[hb]
        \centering
        \caption{List of galaxy clusters hosting a central radio galaxy classified as a GPS source and without evidence for extended radio emission.}
        \label{tab:objects}
        \begin{tabular}{l|c|c|c|c|c|c|c}
                \hline
                 Object & RA & DEC &  z & kpc/$''$ & $\nu_{\text{p}}$ & LLS & $t_{\text{kin}}$\\
                 & & & & & [GHz] & [pc] & [yr]\\
                \hline
                  {\bf RX~J1350+0940} & 13:50:22.1 & +09:40:10.6 & 0.133 & 2.4 & 3.9 & 90 & $7\times10^{2}$ \\
                 \hline
                 Abell 1885 & 14:13:43.73 & +43:39:45.0 & 0.088 & 1.6 & 2.5 & - & -  \\
                 \hline
                 {\bf ClG~J0242-2132} & 02:42:35.9 & -21:32:26 & 0.314 & 4.6 & 0.8 & 200 & $1.6\times10^{3}$\\
                 \hline
                 RX J2341+0018 & 23:41:6.8 & +00:18:34.1 & 0.277 & 4.2 & 0.4 & - & - \\
                 \hline
                 RX J0132-0804 & 01:32:41.1 & -08:04:06 & 0.148 & 2.6 & 0.1 & - & - \\
                 \hline
        \end{tabular}
 \tablefoot{The two objects in bold are the focus of this article. (1) Object name; (2) right ascension; (3) declination; (4) redshift; (5) conversion between physical and angular scales at the object's redshift; (6) rest-frame peak frequency (from \citealt{2015MNRAS.453.1223H}); (7) largest linear size from VLBA data (see \citealt{2020A&A...644A.159C}); and (8) kinematic age of the source (LLS$/2v$) assuming an expansion speed of $v=0.2c$.}
\end{table*}

\section{Methods to derive entropy and cooling time profiles}\label{app:alternative}
\begin{figure*}[h]
   \centering
   \includegraphics[width=\linewidth]{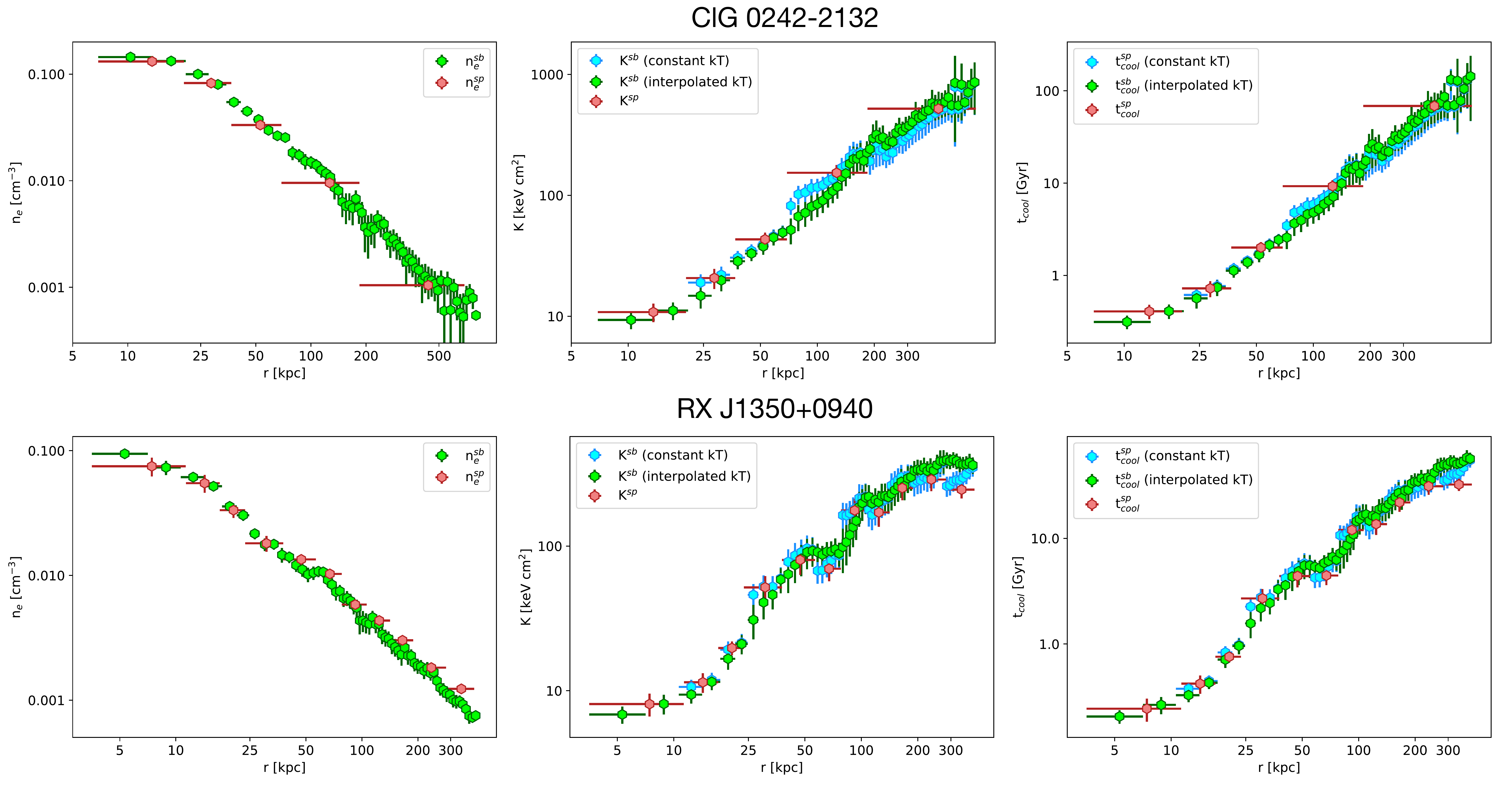}
      \caption{Comparison between the profiles of density, entropy, and cooling time obtained with different methods. Green corresponds to the profiles obtained with the method described in \cref{subsec:profiles} and \cref{app:alternative}; cyan is the result of combining the density profile from the surface brightness profile with the temperature profile (without interpolating the temperature profile); red is the result of combining the density and the temperature derived from spectral fitting. 
              }
         \label{comparison}
\end{figure*}

The profiles shown in Fig. \ref{imgcluster} have been obtained by following a similar procedure to that adopted for the ACCEPT clusters in order to enable a fair comparison. 
In particular, as reported in \cref{subsec:profiles}, the ICM temperature was directly measured from the spectra of concentric annuli with at least 1000 counts per bin. This requirement has resulted in 11 annular rings for RX J1350+0940 (with a size of 3'', 2.5'', 2.8'', 6'', 8'', 9'', 12'', 15'', 20'', 40'', and 50'') and five annular rings for CLG J0242-2132 (with a size of 3.3'', 3.6'', 7'', 25'', and 110''). Instead, the density was determined by deprojecting a background-subtracted, exposure-corrected surface brightness profile extracted in the 0.5-2.0 keV band, with a bin size of 1.5'' (3.5 kpc for RX~J1350+0940 and 7 kpc for CLG~J0242-2132). The deprojected profile was then converted to an electron density profile by providing the count rate and normalization of the spectrum in each annulus of the temperature profile \citep{2011A&A...526A..79E}. The temperature profile was interpolated on the grid of the density profile, and the two quantities were combined to obtain the gas entropy and cooling time.
\\ In the following, we present how different methods of deriving the ICM temperature and density (and, in turn, entropy and cooling time) have returned consistent result.  
As reported above, the temperature of the ICM has been measured from fitting the spectra from concentric annuli centered on the BCG with a deprojected thermal model. The electron density of the ICM can be determined not only by converting a deprojected surface brightness radial profile to a radial density profile, but also from the deprojected normalization of the spectra (see \citealt{2021MNRAS.503.4627U} for a comparison between the two methods).
For the sake of clarity, we label the electron density derived from spectral fitting as $n_{e}^{sp}$ and that obtained from surface brightness analysis $n_{e}^{sb}$. In the left panels of Fig. \ref{comparison}, we show the comparison between the two density profiles for ClG~J0242-2132 and RX~J1350+0940, which are consistent with each other. \\ To obtain the entropy and cooling time of the ICM, it is necessary to combine the temperature and density (Eq. \ref{entro} and Eq. \ref{tcool}). The combination of temperature and $n_{e}^{sp}$ is straightforward, given that the two quantities have been derived from the same radial bins. The combination of temperature and $n_{e}^{sb}$ requires one to associate a temperature measurement to each of the -- more refined -- radial bins of the density profile. The temperature can be either (i) interpolated over the bins of the density profile (as reported in \cref{subsec:profiles}), or (ii) assumed to be constant within the radial range covered by the spectral bins. In the middle and right panels of Fig. \ref{comparison}, we show the comparison between the different methods of deriving the entropy and cooling time radial profiles of the two clusters. The different methods are in good agreement with each other, supporting the results discussed in \cref{subsec:profiles}. 
\\ We note that ClG~J0242-2132 is included in the ACCEPT sample, and it was reported to have a larger central entropy than the one we measured in our work. However, in \citet{2009ApJS..182...12C}, the central point source was not excised, which resulted in the likely inclusion of nonthermal emission in the spectrum of the inner $\sim$30 kpc, shifting the measured temperature to higher values. Indeed, the central temperature reported by \citet{2009ApJS..182...12C} is $\sim$4 keV, which is higher than our measurement (central source excised) of 2.6 keV.

\end{appendix}

\end{document}